\documentclass[aps,prc,twocolumn,groupedaddress,showpacs]{revtex4}

\usepackage{graphicx}

\begin{document}

\title{Density Dependence of the Symmetry Energy and the Equation of State of Isospin Asymmetric Nuclear Matter}
\author{D.V. Shetty, S.J. Yennello, and G.A. Souliotis}
\affiliation{Cyclotron Institute, Texas A$\&$M University, College Station, Texas 77843, USA}
\date{May 10, 2005}

\begin{abstract}
The density dependence of the symmetry energy in the equation of state of isospin asymmetric 
nuclear matter is studied using the isoscaling of the fragment yields and the antisymmetrized 
molecular dynamic calculation. It is observed that the experimental data at low densities are 
consistent with the form of symmetry energy, E$_{sym}$ $\approx$ 31.6 ($\rho/\rho_{\circ})^{0.69}$, 
in close agreement with those predicted by the results of variational many-body calculation. A 
comparison of the present result with those reported recently using the NSCL-MSU data suggests 
that the heavy ion studies favor a dependence of the form, E$_{sym}$ $\approx$ 31.6 ($\rho/\rho_{\circ})^{\gamma}$, 
where $\gamma$ = 0.6 - 1.05. This constrains the form of the density dependence of the symmetry 
energy at higher densities, ruling out an extremely `` stiff " and `` soft " dependences.
\end{abstract}


\maketitle

The Equation Of State (EOS) of isospin asymmetric (N $\neq$ Z) nuclear matter is a fundamental 
quantity that determines the properties of systems as small and light as an atomic nucleus, and as 
large and heavy as a neutron star \cite{DAN02, LAT04, STE05}. The key ingredient in the EOS of 
asymmetric nuclear matter is the density dependence of the symmetry energy. Theoretical 
studies \cite{DIE03, WIR88, LEE98, LIU02, KAI02} based on microscopic many-body calculations 
and phenomenological approaches predict various different forms of the density dependence of the 
symmetry energy. In general, two different forms have been identified \cite{STO03}. One, where the 
symmetry energy increases monotonically with increasing density (`` stiff " dependence) and the other, 
where the symmetry energy increases initially up to normal nuclear density and then decreases at 
higher densities (`` soft " dependence).
\par
Determining the exact form of the density dependence of the symmetry energy is important for studying 
the structure of neutron-rich nuclei \cite{BRO00, HORO01, FUR02, OYA98}, and studies relating to 
astrophysical origin, such as the structure of neutron stars and the dynamics of supernova 
collapse \cite{LAT91, LEE96, LAT01, PET95, LAT00, HIX03}. For example, a `` stiff " density dependence 
of the symmetry energy is predicted to lead to a large neutron skin thickness compared to a `` soft " 
dependence \cite{OYA98, HOR01, HORO01, HOR02}. Similarly, a `` stiff " dependence of the symmetry 
energy can result in rapid cooling of a neutron star, and a larger neutron star radius, compared to a soft 
density dependence \cite{LAT94, SLA02}.
\par
In a heavy ion reaction, the dynamics of the collision between two heavy nuclei is also sensitive to the 
density dependence of the symmetry energy \cite{BAL98,BAR05}. One can therefore carry out 
laboratory-based experiments to constrain this dependence. Recently \cite{ONO03},  the fragment yields 
from heavy ion collisions simulated within the Antisymmetrized Molecular Dynamics (AMD) calculation 
were reported to follow a scaling behavior of the type,
\begin{equation}
     Y_{2}(N,Z)/Y_{1}(N,Z) \propto e^{\alpha N + \beta Z}
\end{equation}
where the parameters $\alpha$ and $\beta$ are related to the neutron-proton content of the fragmenting 
source, and $Y_{1}$ and $Y_{2}$ are the yields from two different reactions. A linear relation between the 
isoscaling parameter $\alpha$, and the difference in the isospin asymmetry ($Z/A)^{2}$ of the fragments, 
with appreciably different slopes, was predicted for two different forms of the density dependence of the 
symmetry energy ; a `` stiff " dependence (obtained from Gogny-AS interaction) and a `` soft " dependence 
(obtained from Gogny interaction).  
\par
In this work, we show that the experimentally measured scaling parameter $\alpha$, favors a stiff density 
dependence of the symmetry energy, i.e. Gogny-AS interaction, and can be parametrized as 
E$_{sym}$ $\approx$ 31.6 ($\rho/\rho_{\circ})^{\gamma}$, where $\gamma$ = 0.69. The present observation 
is consistent with the EOS of Akmal and Pandharipande obtained from the many-body variational 
calculations \cite{AKM97, AKM98}.  
\par
The measurements were carried out at the Cyclotron Institute, Texas A$\&$M University using 
beams of $^{40}$Ar, $^{40}$Ca, $^{58}$Fe and $^{58}$Ni from the K500 Superconducting 
Cyclotron on $^{58}$Fe and $^{58}$Ni targets at 25, 30, 33, 40, 45, 47 and 53 MeV/nucleon. 
Details of the experimental measurements and analysis can be found in Ref. \cite{SHE04}. 
\par
Figure 1, shows the experimentally determined isoscaling parameter $\alpha$, obtained from 
the fragment yields as a function of the beam energy. The different symbols correspond to various 
combinations of the reactions chosen for extracting the isoscaling parameters. The solid and the 
dotted lines are the exponential fits to the data.
    \begin{figure}
    \includegraphics[width=0.5\textwidth,height=0.45\textheight]{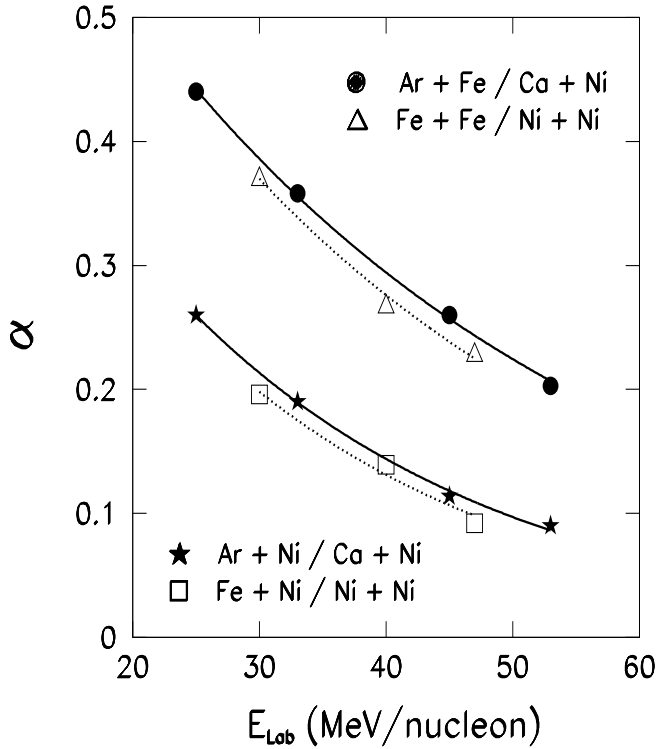} 
    \caption{Experimental isoscaling parameter $\alpha$, as a function of the beam energy. The solid 
             circles are from the Ar + Fe and Ca + Ni reactions. The open triangles are from Fe + Fe and 
             Ni + Ni reactions. The solid stars are from Ar + Ni and Ca + Ni reactions. The open squares 
                are from Fe + Ni and Ni + Ni reactions.}
   \end{figure}
\par
As mentioned earlier and shown in \cite{ONO03}, the parameter $\alpha$, is related to the difference 
in the fragment isospin asymmetry $(Z/A)^{2}$, through a linear relation of the form

\begin{equation}
      \alpha = \frac{4C_{sym}}{T} {[(Z/A)_{1}^{2} - (Z/A)_{2}^{2}]}
\end{equation}

where $C_{sym}$ is the symmetry energy and $T$ is the temperature at which the fragments are 
formed. The quantity $(Z/A)_{1}^{2} - (Z/A)_{2}^{2}$, is the difference in the isospin asymmetry 
of the fragments in the two reaction systems. For the present systems, the isospin asymmetry of 
the fragments were evaluated at ${\it {t}}$ = 300 fm/{\it {c}} of the dynamical evolution from the 
AMD calculations as discussed extensively in Ref. \cite{SHE04}. 
\par
Fig. 2 shows  the $\alpha$ parameters plotted as a function of the difference in the fragment 
asymmetry for the beam energy of 35 MeV/nucleon. The solid and the dotted lines are the AMD 
predictions using the `` soft " (Gogny) and the `` stiff " (Gogny-AS) density dependence of the 
symmetry energy, respectively.
    \begin{figure}
    \includegraphics[width=0.5\textwidth,height=0.45\textheight]{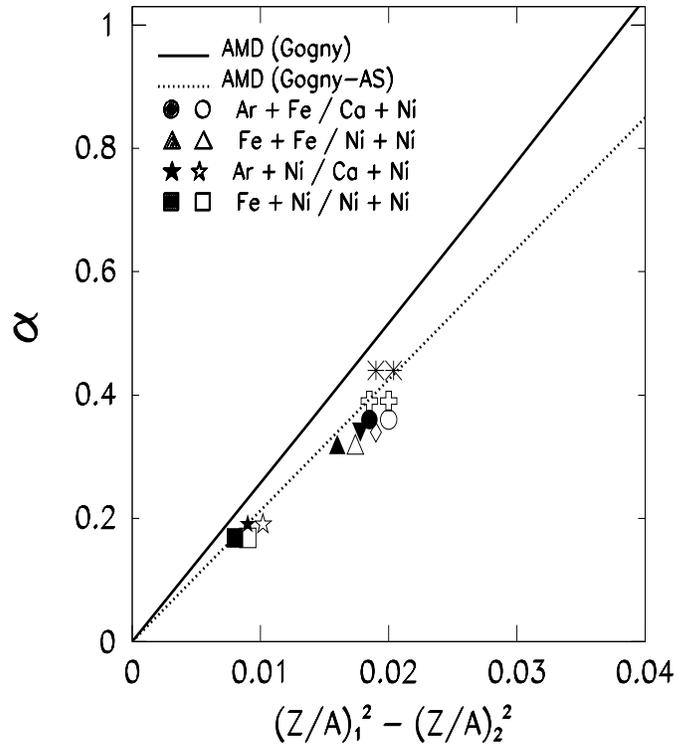} 
    \caption{Scaling parameter $\alpha$, as a function of the difference in fragment asymmetry for 
                 35 MeV/nucleon. The solid and the dotted lines are the AMD calculations for the Gogny 
                 and Gogny-AS interactions, respectively \cite{ONO03}. The solid and the hollow, squares, 
                 stars, triangles and circles are from the present work as described in the text. The other symbols 
                 corresponds to data taken from \cite{GER04} (asterisks) and \cite{BOT02} (crosses, diamonds, 
                 inverted triangles).} 
   \end{figure}
The solid and the hollow symbols (squares, stars, triangles and circles) are the results of  the present 
measurements for the two different values of the fragment asymmetry, assuming Gogny and Gogny-AS 
interactions, respectively. Also shown in the figure are the scaling parameters (asterisks, crosses, diamond 
and inverted triangle) taken from various other works \cite{GER04, BOT02} in the literature. One observes 
from the figure that the experimentally determined $\alpha$ parameter increases linearly with increasing 
difference in the asymmetry of the two systems as predicted by the AMD calculation. Also the data points 
are in closer agreement with those predicted by the Gogny-AS interaction (dotted line) than those from the 
usual Gogny force (solid line). 
\par
It should be mentioned that in the above comparison between the data and the calculation, the
corrections for the isoscaling parameter $\alpha$ due to the secondary de-excitation of the fragments are not 
taken into account. The slightly lower values 
of the isoscaling parameters (symbols) from the present measurements with respect to the Gogny-AS values (dotted line) 
could be due to the small secondary de-excitation effect of the fragments not accounted for in this comparison.
It has been reported by Ono {\it {et al.}} \cite{ONO05}, that the sequential decay effect in the dynamical calculations
can affect the $\alpha$ value by as much as 50 $\%$. On the other hand, dynamical calculation carried out
by Tian {\it {et al.,}} \cite{TIA06} for the same systems and energy as studied by Ono {\it {et al.}}, using Isospin
Quantum Molecular Dynamic (IQMD) model, shown no significant difference between the primary and the 
secondary $\alpha$. 
\par
Due to the large discrepancy that exist in the determination of the primary fragment excitation energy from the 
dynamical model calculations, it is difficult to estimate the effect of secondary de-excitation in dynamical models 
at this moment \cite{SHET06}. We have therefore assumed the effect of the sequential decay to be negligible, in the 
above comparison. A small correction of about 10 - 15 $\%$, as determined from various statistical model studies
\cite{TSAN01}, results in a slight increase in the $\alpha$ values bringing them even closer to the dotted line. 
Note the asterisks symbols shown in the figure, and taken from the Ref. \cite{GER04}, has already been corrected. 
The closer agreement of the experimental data with the Gogny-AS type of interaction, therefore, appears
to suggest a stiffer density dependence of the symmetry energy rather than the soft Gogny interaction. 
\par
Recently, Chen {\it {et al.}} \cite{CHE05} also showed, using the isospin dependent Boltzmann-Uehling-Uhlenbeck 
(IBUU04) transport model calculation, that a stiff density dependence of the symmetry energy 
parametrized as, E$_{sym}$ $\approx$ 31.6 ($\rho/\rho_{\circ})^{1.05}$ explains well the isospin 
diffusion data \cite{TSA04} from NSCL-MSU (National Superconducting Cyclotron Laboratory at Michigan 
State University).  Their  calculation was also based on a momentum-dependent Gogny effective 
interaction. However, the present measurements on isoscaling gives a slightly softer density dependence 
of the symmetry energy at higher densities than those obtained by Chen {\it {et al.}}
\par
This is clear from figure 3, which shows the parameterization of various theoretical predictions of the 
density dependence of the nuclear symmetry energy in isospin asymmetric nuclear matter. The 
dot-dashed, dotted and the dashed curve corresponds to those from the momentum dependent 
Gogny interactions used by Chen {\it {et al.}} to explain the isospin diffusion data. These are given 
as, E$_{sym}$ $\approx$ 31.6 ($\rho/\rho_{\circ})^{\gamma}$, where, $\gamma$ = 1.6, 1.05 and 0.69, 
respectively. The solid curves and the solid points corresponds to those from the Gogny and 
Gogny-AS interactions used to compare with the present isoscaling data. As shown by Chen {\it {et al.}}, 
the dependence parameterized by E$_{sym}$ $\approx$ 31.6 ($\rho/\rho_{\circ})^{1.05}$ (dotted curve)
explains the NSCL-MSU data on isospin diffusion quite well. On the other hand, the isoscaling data
from the present work can be explained well by the Gogny-AS interaction (solid points).
    \begin{figure}
    \includegraphics[width=0.5\textwidth,height=0.45\textheight]{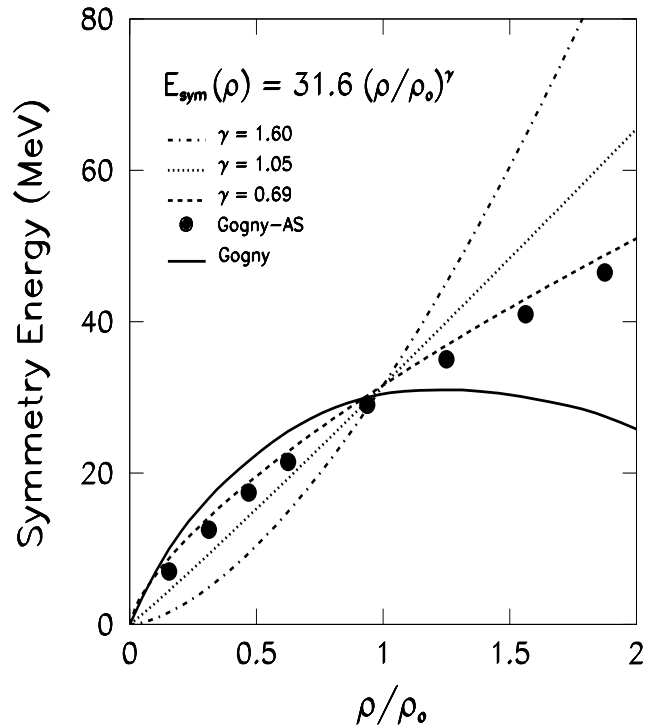}
    \caption{Parameterization of the various forms of the density dependence of the nuclear symmetry energy 
                used in the analysis of the present measurements on isoscaling data and the isospin diffusion 
                measurements of NSCL-MSU \cite{CHE05}. The various curves are as described in the text.} 
   \end{figure}
Both measurements yield similar results at low densities with significant difference at higher densities. 
It is interesting to note that by parameterizing the density dependence of the symmetry energy that 
explains the present isoscaling data, one gets, E$_{sym}$ $\approx$ 31.6 ($\rho/\rho_{\circ})^{\gamma}$, 
where $\gamma$ = 0.69. This form of the density dependence of the symmetry energy is consistent 
with the parameterization adopted by Heiselberg and Hjorth-Jensen in their studies on neutron stars \cite{HEI00}. 
By fitting earlier predictions of the variational calculations by Akmal {\it {et al.}} \cite{AKM97, AKM98}, 
where the many-body and special relativistic corrections are progressively incorporated, Heiselberg 
and Hjorth-Jensen obtained a value of E$_{sym}$($\rho_{\circ}$) = 32 MeV and $\gamma$ = 0.6, 
similar to those obtained from the present measurements. The present form of the density dependence is 
also consistent with the findings of Khoa {\it {et al.}} \cite{KHO05}, where a comparison of the 
experimental cross-sections in a charge-exchange reaction with the Hartree-Fock calculation 
using the CDM3Y6 interaction \cite{KHO97}, reproduces well the empirical half-density point of 
the symmetry energy obtained from the present work (see fig. 2 of Ref. \cite{KHO05}).
\par
The observed difference in the form of the density dependence of the symmetry energy between the 
present measurement and those obtained by Chen  {\it {et al.}} is not surprising. Both measurements 
probe the low density part of the symmetry energy and are thus less sensitive to the high density 
region. But the important point to be noted is that both measurements clearly favor a stiff density 
dependence of the symmetry energy at higher densities, ruling out the very `` stiff '' 
(dot-dashed curve) and very `` soft '' (solid curve) predictions. These results can thus be used 
to constrain the form of the density dependence of the symmetry energy at supranormal densities 
relevant for the neutron star studies.
\par
It should be mentioned that the calculations in both the above described works assume a similar 
value for the symmetry energy at normal nuclear density (about 31 MeV). Although, numerous 
many-body calculations \cite{ZUO99, BRA85, PEA00, DIE03} and those from the empirical liquid 
drop mass formula \cite{MYE66, POM03} predict symmetry energy near normal nuclear density 
to be around 30 MeV, a direct experimental determination of the symmetry energy does not exist. 
\par
Recently, Khoa {\it {et al.}} \cite{KHO05}, analyzed the experimental cross-section data \cite{COR98, COR97} 
using the isospin dependent CDM3Y6 interaction of the optical potential in a charge 
exchange $p(^{6}He, ^{6}Li^{*})n$ reaction. Their analysis probed mainly the surface part of 
the form factor and hence appropriate for densities close to the normal nuclear density. Based on 
their results and the Hartree-Fock calculation of asymmetric nuclear matter using the same effective 
nucleon-nucleon interaction, they estimate the most realistic value of the symmetry energy to be 
about 31 MeV. An accurate determination of the neutron skin thickness $\Delta R$, from the 
parity-violating electron scattering measurement \cite{HOR01} is however, likely to provide a 
more precise determination of the symmetry energy near normal nuclear density.
\par
In view of the findings from the present measurements and those of Chen {\it {et al.}}, we believe 
that the best estimate of the density dependence of the symmetry energy that can be presently 
extracted from heavy ion reaction studies is, E$_{sym}$ $\approx$ 31.6 ($\rho/\rho_{\circ})^{\gamma}$, 
where $\gamma$ = 0.6 - 1.05. It must be mentioned that the present comparison between the experimental 
data and the theoretical calculation is model dependent. Any 
modification to the compressibility in the equation of state could affect the pressure at sub-saturation 
densities and thus the agreement between the data and the calculation. 
\par
This work was supported in part by the Robert A. Welch Foundation through grant No. A-1266, 
and the Department of Energy through grant No. DE-FG03-93ER40773. 
\  \\
\  \\
{\bf Note :} Several other authors have now reported similar conclusions using other observables since this 
article was first submitted.

\end{document}